\documentclass[aps,prb,showpacs,twocolumn,floats,epsfig,pdflatex, superscriptaddress]{revtex4-2}
\usepackage[T1]{fontenc}
\usepackage[latin9]{inputenc}
\usepackage{color}
\definecolor{shadecolor}{rgb}{1, 0, 0}
\usepackage{verbatim}
\usepackage{float}
\usepackage{framed}
\usepackage{amsmath}
\usepackage{amssymb}
\usepackage{graphicx}

\usepackage{slashed}

\newcommand\e{\textrm{e}}
\newcommand\di{\textrm{d}}

\makeatletter
\PassOptionsToPackage{caption=false}{subfig}
\usepackage{hyperref}
\hypersetup{
breaklinks=true,
colorlinks=true,
citecolor=blue,
linkcolor=blue,
filecolor=blue,
urlcolor=blue
}
\IfFileExists{lmodern.sty}{\usepackage{lmodern}}{}

\makeatother

\usepackage{babel}
\usepackage{braket}

\begin{document}

\title{Phononic dynamical axion in magnetic Dirac insulators}

\author{M. Nabil Y. Lhachemi}
\affiliation{D\'epartement de physique, Institut quantique and Regroupement Qu\'eb\'ecois sur les Mat\'eriaux de Pointe,  Universit\'{e} de Sherbrooke, Sherbrooke, Qu\'{e}bec J1K 2R1, Canada}
\affiliation{Department of physics and Astronomy, Stony Brook University, Stony Brook, NY 11794, USA}
\author{Ion Garate}
\affiliation{D\'epartement de physique, Institut quantique and Regroupement Qu\'eb\'ecois sur les Mat\'eriaux de Pointe,  Universit\'{e} de Sherbrooke, Sherbrooke, Qu\'{e}bec J1K 2R1, Canada}

\date{\today}                                       

\begin{abstract}
In cosmology, the axion is a hypothetical particle that is currently considered as candidate for dark matter. 
In condensed matter, a counterpart of the axion  (the ``axion quasiparticle'') has been predicted to emerge in magnetoelectric insulators with fluctuating magnetic order and in charge-ordered Weyl semimetals.
To date, both the cosmological and condensed-matter axions remain experimentally elusive or unconfirmed.
Here, we show theoretically that ordinary lattice vibrations can form an axion quasiparticle in Dirac insulators with broken time- and space-inversion symmetries, even in the absence of magnetic fluctuations. 
The physical manifestation of the phononic axion is a magnetic-field-induced phonon effective charge, which can be probed in optical spectroscopy.
By replacing magnetic fluctuations with lattice vibrations, our theory widens the scope for the observability of the axion quasiparticle in condensed matter.

 \end{abstract}
\maketitle

\section{Introduction}

In the past 15 years, much effort has been devoted in condensed matter physics to the study 
of the ``axion term''  $\theta {\bf E}\cdot{\bf B}$, where ${\bf E}$ and ${\bf B}$ are the electric and magnetic fields (respectively) and  $\theta$ is the so-called axion field \cite{nenno2020axion, sekine2021axion}.  
Such axion term arises in the electromagnetic Lagrangian of certain conventional magnetoelectric materials \cite{vanderbilt2018berry}, Dirac insulators \cite{qi2008topological}, Weyl semimetals \cite{wang2013chiral} and superconductors with vortices \cite{qi2013axion}.
 
In 2010, a seminal paper~\cite{Li2010} proposed that magnetic topological insulators with broken space-inversion  and time-reversal symmetries host a time-dependent $\theta$, whose dynamics originates from the fluctuations of the magnetic order parameter. 
The action describing magnetic fluctuations could thus be mapped to that of the cosmological axion.
This ``axion quasiparticle'' of condensed matter was predicted to produce exotic physical effects \cite{Li2010,ooguri2012instability,sekine2016chiral} and has been recently recognized as a  detection tool for the cosmological axion \cite{schutte2021axion}.
Unfortunately, the axion quasiparticle remains undetected in insulators, in part because 
the dynamical $\theta$ induced by magnetic fluctuations is rather small \cite{chaoxing2020making}.
On a related note, an axion quasiparticle was predicted to emerge in Weyl semimetals with charge density wave order \cite{wang2013chiral, li2023axionic}. While evidence supporting the prediction of Ref.~\cite{wang2013chiral} has been reported in recent magnetotransport measurements \cite{gooth2019axionic}, such claim has not been confirmed in subsequent experiments \cite{cohn2020magnetoresistance, sinchenko2022does} and the properties of the putative axion quasiparticle remain to be studied in detail \cite{mottola2023axions}.


Ever since the publication of Ref.~\cite{Li2010},  magnetic fluctuations and their coupling to electrons have been widely regarded as quintessential for the emergence of the axion quasiparticle in magnetic insulators.
In this paper, we propose an alternate route for the generation and detection of the axion quasiparticle, by showing 
that ordinary lattice vibrations in a Dirac insulator with broken inversion and time-reversal symmetries can induce a sizeable dynamical axion term in the absence of magnetic fluctuations.

The rest of this work is organized as follows. In Sec.~\ref{sec:PEC}, we review the definition of the phonon effective charge within the action formalism, and notice that magnetic-field-induced phonon effective charge is a signature of a phononic axion. Then, we identify a basic mechanism whereby lattice vibrations producing fluctuations in the scalar Dirac mass lead to an axion term in the electromagnetic Lagrangian.

In Sec.~\ref{sec:micro}, we use a functional integral formalism to calculate the phonon-induced axion term in three dimensional Dirac insulators with broken time- and space-inversion symmetries. 
We study a model Hamiltonian of a Dirac fermion coupled to electromagnetic gauge fields and to a phonon field. 
By integrating out the Dirac fermion, we find an axion term $\delta\theta_{\rm ph} {\bf E}\cdot{\bf B}$, where $\delta\theta_{\rm ph}$ is proportional to the lattice displacement from equilibrium and to the electron-phonon coupling.
We calculate $\delta\theta_{\rm ph}$ as a function of the phonon frequency and the temperature, identifying a maximum of $\delta\theta_{\rm ph}$ when the phonon frequency coincides with the energy gap of the insulator.
While lattice-mediated magnetoelectric responses have been evaluated in earlier density functional theory studies \cite{iniguez2008first, malashevich2012full}, to our knowledge there has been neither a calculation of the dynamics of a phonon-induced axion term, nor an investigation of the role of electron-phonon interactions. Both items are central to our paper.

In Sec.~\ref{sec:exp}, we point out the experimental signatures of the phononic axion: the magnetic-field-induced phonon effective charge is in principle measurable in infrared or Raman spectroscopies.
We conclude the paper in Sec.~\ref{sec:disc} with a summary and discussion.

\section{Phonon effective charge}
\label{sec:PEC}

\subsection{Definition}

The central quantity of the present paper is the phonon effective charge (PEC). It is defined as the change (in linear response approximation) of the unit cell dipole moment produced by a given phonon mode~\cite{keldysh, gonze1997}:
\begin{equation}\label{eq:Qdef}
    \textbf{Q}(q_0,{\bf q})\equiv {\cal V}_c\frac{\partial\textbf{P}(q_0,{\bf q})}{\partial u(q_0,{\bf q})}\bigg\rvert_{u=0}, 
\end{equation}
where $q_0$ and ${\bf q}$ are the phonon frequency and wave vector, respectively, $u$ is the phonon normal coordinate, ${\cal V}_c$ is the unit cell volume and ${\bf P}$ is the electric polarization.
For brevity, we omit the phonon mode index throughout this work.

Because ${\bf P}$ is a polar vector, the only phonon modes with ${\bf Q}\neq 0$ are those that transform as polar vectors under the symmetry operations of the crystal~\cite{anastassakis1980morphic}.
Long-wavelength acoustic phonons have negligible PEC in spite of transforming like polar vectors, as they describe an overall translation of the unit cell (without producing a change in the intraunit cell dipole moment). 
In this work, we will focus on long-wavelength optical phonons, which can have a sizable magnitude of ${\bf Q}$ (of the order of an electron charge) and are detectable in optical spectroscopy.


\subsection{Connection with the action formalism}

In the imaginary-time action formalism, 
the PEC appears as a direct phonon-photon coupling~\cite{Rinkel2017,Rinkel2019}:
\begin{align}\label{eq : action Q0}
    S_{\textbf{Q}}[A, u] &= \frac{i}{\hbar\beta \cal V}\sum_{q}  \textbf{P}_{\rm ph}(q)\cdot\textbf{E}(-q)\nonumber\\
     &=  \frac{i}{\hbar\beta \cal V}\frac{1}{{\cal V}_c}\sum_{q}  \textbf{Q}(q)\cdot\textbf{E}(-q)u(q),
\end{align}
where ${\cal V}$ is the crystal volume, $q$ is the shorthand notation for wave vector and (imaginary) frequency,
$\beta=1/(k_B T)$ is the inverse of temperature, ${\bf P}_{\rm ph}$ is the phonon contribution to electric polarization, and ${\bf E}$ is the electric field.
For Fourier transforms, we use the conventions $f(x) =1/(\beta {\cal V} ) \sum_q \exp(-i q\cdot x) f(q)$
and $f(q) = \int d x \exp(i q\cdot x) f(x)$, where $x$ is the shorthand notation for position and (imaginary) time. 




\subsection{Connection with the dynamical axion}


The action describing the electrodynamics of certain three dimensional Dirac insulators contains, aside from the conventional Maxwell term,  a topological term \cite{sekine2021axion}
\begin{equation}\label{eq : theta action}
    S_{\theta}[A] =  \frac{i e^2}{4\pi^2\hbar}\int \di x\, \theta (x)\textbf{E}(x)\cdot\textbf{B}(x),
\end{equation}
where ${\bf B}$ is the magnetic field, $\theta$ is the axion angle and $e$ is the electron's charge.
In insulators with time-reversal or inversion symmetry, $\theta$ is a quantized topological invariant, with $\theta = 0 \,({\rm mod}\, 2\pi)$ for trivial insulators and $\theta = \pi \, ({\rm mod}\, 2\pi)$ for topological insulators. 
When those symmetries are broken, $\theta=\theta(x)$ becomes a function of spacetime known as the dynamical axion.

If the magnetic field ${\bf B}$ is constant in time and uniform in space (${\bf B}={\bf B}_0$), Eq.~(\ref{eq : theta action}) can be rewritten as
\begin{equation}\label{eq : theta action2}
    S_{\theta}[A] =\frac{i e^2}{4\pi^2\hbar} \frac{1}{\hbar\beta \cal V}\sum_q \theta (q)\textbf{B}_0\cdot\textbf{E}(-q).
\end{equation}
Comparing Eqs.~(\ref{eq : theta action2}) and (\ref{eq : action Q0}), we extract an axion-related PEC of the form 
\begin{equation}
\label{eq:Qtheta}
{\bf Q}_\theta(q) =  {\cal V}_c\frac{e^2}{4\pi^2\hbar} \frac{\partial\theta(q)}{\partial u(q)}\bigg\rvert_{u=0} {\bf B}_0,
\end{equation}
where we have assumed that $\theta$ is an analytic function of the lattice displacement.
Accordingly, the PEC in Dirac insulators consists of two distinct terms:
\begin{equation*}
    \textbf{Q} = \textbf{Q}_0 + \textbf{Q}_{\theta}.
\end{equation*}
The first term is the conventional PEC coming from the Maxwell action. The second term is a topological or axionic PEC, since it comes from $S_\theta$. From Eqs.~(\ref{eq:Qdef}) and (\ref{eq : theta action2}), we infer
\begin{equation}\label{eq : PEC topo}
    \textbf{Q}_{\theta}= {\cal V}_c\frac{\partial \textbf{P}^{\theta}}{\partial u}\bigg\rvert_{u=0},
\end{equation}
where $\textbf{P}^{\theta}=-i\hbar \beta {\cal V} (\delta S_\theta/\delta {\bf E})$ is the electric polarization induced by a magnetic field \cite{Qi2008}. 
Thus, ${\bf Q}_\theta$ can be interpreted as the phonon-induced modulation of the topological magnetoelectric effect. It manifests itself if and only if $\theta$ depends on the phonon field $u$.
The theoretical \cite{Rinkel2017, Rinkel2019} and experimental \cite{Yuan2020} existence of ${\bf Q}_\theta$ was reported in Weyl semimetals, i.e. Dirac materials characterized by zero Dirac mass. We will see in the next sections that Dirac insulators can also have this nontrivial PEC.


\subsection{Physical origin of ${\bf Q}_\theta$}
\label{sec:PECax}
 
We now discuss a simple mechanism through which $\theta$ can depend on $u$.
Let us consider a Dirac insulator with time-reversal and space-inversion symmetries, where $\theta$ is quantized. 
The low-energy electronic bands are described by a Dirac Hamiltonian with a mass.
The sign of that mass determines whether $\theta$ is equal to $0$ or $\pi$ (mod $2\pi$)  \cite{qi2008topological, rosenberg2010}; thus, changing the sign of mass (i.e. going through a band inversion) implies a topological phase transition.


In the presence of a long-wavelength (${\bf q}\simeq 0$) optical phonon coupled to electrons, the Dirac mass is modulated
\footnote{This situation is fundamentally different from the case of topological semimetals, where a fluctuating energy gap can be produced only for relatively large phonon wave vectors connecting two Weyl nodes. }
 as a function of time $t$:
\begin{equation}\label{eq : Dirac mass}
    m(t) = m + \delta m(t) = m + g u(t),
\end{equation}
where $m$ is the unperturbed Dirac mass, $\delta m(t)$ is the instantaneous mass fluctuation induced by electron-phonon interactions,  and $g$ is the {\em difference} between the optical deformation potentials for the bottom of the conduction band and the top of the valence band. 
Because generic Dirac materials do not have electron-hole symmetry, the optical deformation potentials are indeed different for the conduction band minimum and the valence band maximum. Accordingly, $g$ is generically nonzero.
In addition, because the Dirac mass respects all crystal symmetries, a fully symmetric $A_1$ phonon can lead to $g\neq 0$.
Moreover, we anticipate that $g$ is independent of $m$ for small $m$, so that phonons can invert the bandgap.

If $ |g u|<|m|$, the phonon-induced modulations of $m(t)$ are not large enough to close the energy gap of the insulator; the (quantized) value of $\theta$ is thus independent of $u$. In contrast, if $|g u|>|m|$, phonons produce dynamical (oscillatory) band inversions and therefore change the value of $\theta$.
We infer that $\theta$ depends on $u$ when $|m| < |g u|$. 
Since $|g u|$ is a small energy scale, it follows that $\theta$ depends on $u$ (and thus ${\bf Q}_\theta$ is nonzero) only in the vicinity of a topological phase transition ($m\simeq 0$).  
As such, ${\bf Q}_\theta$ is a diagnostic tool for the topological phase transition. 
This idea can be regarded as a counterpart of Ref.~[\onlinecite{yu2020piezoelectricity}] for three dimensional systems.

A caveat for the preceding idea is in order.  This caveat is analogous to the one that applies to the surface Hall effect originated from the axion term \cite{qi2008topological}.
If the entire crystal has time-reversal symmetry, there is no way to determine how $\theta$  changes in the course of a band inversion; e.g. it could equally likely be $0\to \pi$ or $0 \to -\pi$. Yet, these two options correspond to opposite values of ${\bf Q}_\theta$. Thus, one cannot observe any ${\bf Q}_\theta$ when time-reversal symmetry is preserved everywhere.
 In order to have an observable ${\bf Q}_\theta$, time-reversal symmetry must be broken at least on the surface of the material \cite{armitage2019matter}. The symmetry breaking perturbation then dictates the change of $\theta$ across the band inversion (e.g. it chooses between $0\to \pi$ and $0 \to -\pi$).

Below, we will consider the situation where time-reversal symmetry and space-inversion symmetry are broken in the entire bulk. As a consequence, $\theta$ will not be quantized and $\theta$ will depend not only on the sign of $m$, but also on its magnitude. As such, phonons that modulate $m$ will also modulate $\theta$. Unlike in the time-reversal symmetric case, the $u-$dependence of $\theta$ will be nonzero even far from a band inversion. In certain regimes of physical relevance, we will still predict a significant maximum of $|{\bf Q}_\theta|$ as $m$ crosses zero.  Yet, as we will show, a more generic statement is that $|{\bf Q}_\theta|$ has a maximum when the phonon frequency matches the energy gap of the insulator.



\section{Microscopic theory of the phonon-induced axion term}
\label{sec:micro}

In this section, we provide an explicit calculation of $\partial\theta/\partial u$ (and therefore ${\bf Q}_\theta$) for a three dimensional Dirac insulator with broken time-reversal and space-inversion symmetries. 
We take $\hbar\equiv 1$ unless otherwise noted.

\subsection{Partition function}



The starting point is the partition function of the system in the imaginary time formalism, given by
\begin{equation}\label{eq : partition function definition}
    \mathcal{Z} = \int\mathcal{D}[\psi, \Bar{\psi}, A, n, u]\e^{-S},
\end{equation}
where $\psi$ and $\Bar{\psi}$ are fermion Grassmann fields, $n$ is the field associated to amplitude fluctuations of the magnetic order, $u$ is the phonon field and $A$ is the electromagnetic gauge field. 
In Eq.~(\ref{eq : partition function definition}), the total action reads
\begin{equation}\label{eq : S tot 1}
\begin{split}
    S &= S_{\text{ph}}^{(0)}[u] + S_{\text{mag}}^{(0)}[n] + S_{\text{em}}^{(0)}[A] + S_{\textbf{Q}_{0}}[A, u] \\
    &\quad+S_{\text{e}}[\psi, \Bar{\psi}] + S_{\text{int}}[\psi, \Bar{\psi}, A, u, n],
\end{split}
\end{equation}
where the three first terms represent the actions for free phonons, magnetic fluctuations and photons, respectively.
The action $S_{\textbf{Q}_{0}}[A, u]$ is given by Eq.~(\ref{eq : action Q0}), where ${\bf Q}_0$ excludes the contribution from low-energy Dirac fermions (to be computed below).
The action $S_{\e}$ represents free, low-energy electrons of the Dirac material,
\begin{equation}\label{eq : Se pauli}
    S_e[\psi,\Bar{\psi}]=  \int \di x\Bar{\psi}\left(\tau^x\partial_\tau + h_0  \right)\psi,
\end{equation}
where $\partial_\tau$ is the (imaginary) time-derivative,
\begin{equation}\label{eq : hamitonien}
    h_0 = -i  \tau^x\tau^z\boldsymbol{\sigma}\cdot\boldsymbol{\partial} + m + m_5\tau^x\tau^y
\end{equation}
is the Dirac Hamiltonian with the Dirac velocity defined as unity, $\sigma^i$ and $\tau^i$ are the Pauli matrices associated with spin and orbital degrees of freedom, respectively, $m$ is the scalar (symmetry-preserving) mass and $m_5$ is the axial mass. 
The latter breaks space-inversion $\mathcal{P}$ and time-reversal $\mathcal{T}$ within the volume of the insulator, while preserving $\mathcal{P}\mathcal{T}$. 
The microscopic origin of $m_5$ is magnetic (usually antiferromagnetic) order.

The Hamiltonian in Eq.~(\ref{eq : hamitonien}) describes the low-energy electronic bands of a Dirac insulator 
with an energy spectrum 
\begin{equation}
    E_{\pm}(\textbf{k})  =\pm\sqrt{{\bf k }^2 + m^2 + m_5^2}
    \equiv  \pm E_{\textbf{k}}
\end{equation}
and an energy gap 
\begin{equation}
\Delta = 2\sqrt{m^2 + m_5^2}.
\end{equation}
Each energy band is twofold degenerate. 
With Eq.~(\ref{eq : hamitonien}), 
$S_{\e}$ can be rewritten as
\begin{equation}\label{eq : Se Dirac}
    S_e[\psi,\Bar{\psi}] =  \int \di x\Bar{\psi}\left( -i\slashed{\partial}  + m + im_5\gamma^5 \right)\psi,
\end{equation}
where the Feynman slash notation $\slashed{\partial} = \gamma^{\mu}\partial_{\mu} = \boldsymbol{\gamma}\cdot\boldsymbol{\partial} + \gamma^4\partial_{\tau}$ has been used. The matrices $\gamma^{\mu}$, $\mu = 1, 2, 3, 4$, form a representation of the Clifford algebra defined by $\{\gamma_{\mu}, \gamma_{\nu}\} = 2g_{\mu\nu}= -2\delta_{\mu\nu}$. The fifth matrix $\gamma^5 = \gamma^4\gamma^1\gamma^2\gamma^3$, also called chiral matrix, anticommutes with all the other matrices: $\{\gamma^5, \gamma^{\mu}\} = 0$ \cite{Fujikawa}. The form of these matrices in terms of the Pauli matrices can be deduced by comparing Eq.~(\ref{eq : Se pauli}) and Eq.~(\ref{eq : Se Dirac}): $\boldsymbol{\gamma} = -i\tau^y\boldsymbol{\sigma}$, $\gamma^4 = i\tau^x$ and $\gamma^5 = \tau^z$.

Finally, the interacting part of the action in Eq.~(\ref{eq : S tot 1}) reads
\begin{align}\label{eq : action int}
&S_{\text{int}}[\psi, \Bar{\psi}, A, u, n]\notag\\
&= \int \di x\Bar{\psi}\left(-e\slashed{A} + \delta m(u)+i\delta m_5(n)\gamma^5 \right)\psi.
\end{align}
The first term in the right hand side of Eq.~(\ref{eq : action int}) represents the electron-photon interaction which is obtained by minimal coupling. 
The last two terms of Eq.~(\ref{eq : action int}) represent a scalar and pseudosacalar Yukawa interaction, respectively.
In the second term, the electron-phonon interaction leads to a correction of the Dirac mass, i.e. $\delta m (u)= g u$.
Analogously, magnetic fluctuations couple to electrons through a correction of the axial mass, i.e. $\delta m_5(u)\propto n$ \cite{Li2010}. 

An interaction similar to Eq.~(\ref{eq : action int}) has been considered in Ref.~\cite{imaeda2019axion}. In that paper, $\delta m$ originates from the fluctuations of an unspecified order parameter. In our model, $\delta m$ is a consequence of simple lattice vibrations not requiring an order parameter.

\subsection{Dirac fermion contribution to the PEC}

In order to obtain the low-energy electronic contribution to the phonon effective charge, which will result in ${\bf Q}_\theta$, we proceed by integrating out the Dirac fermions. 
Then, the partition function becomes
\begin{equation}
    \mathcal{Z} = \int\mathcal{D}[A, n, u]\e^{-S^{\text{eff}}},
\end{equation}
where 
\begin{equation}
\begin{split}
    S^{\text{eff}} &= S_{\text{ph}}^{(0)}[u] + S_{\text{m}}^{(0)}[n] + S_{\text{em}}^{(0)}[A] + S_{\textbf{Q}_{0}}[A, u] \\ &\quad- \text{Tr} \ln\left[\beta\left(G_0^{-1} + V\right) \right]
\end{split}
\end{equation}
is the effective action for phonons, magnons and photons. In the last term of $S^{\text{eff}}$, the trace Tr is over spacetime as well as over the spin and orbital degrees of freedom, and
\begin{align}
    G_0^{-1} &= -i\slashed{\partial} + m + im_5\gamma^5 \label{eq : Green real}\\
    V(x) &= - e\slashed{A}(x) + \delta m(x) + i\delta m_5(x)\gamma^5 \label{eq : Perturbation real}
\end{align}
are the inverse of the free fermion Green's function and the perturbation, respectively. 
Expanding the last term of $S^{\rm eff}$  in powers of $V$, one gets \cite{Nagaosa, Coleman}
\begin{align}
    S^{\text{eff}} &= S_{\text{ph}}^{(0)}[u] + S_{\text{m}}^{(0)}[n] + S_{\text{em}}^{(0)}[A] + S_{\textbf{Q}_{0}}[A, u] \notag\\
    &- \text{Tr}\ln\left[\beta G_0^{-1}\right] - \sum_{j=1}^{\infty}S_j[A, u, n],
\end{align}
where $S_j = -\text{Tr}\left[-G_0V\right]^j/j$. We will now focus our attention on $S_3$, since it is the lowest-order correction that can host a term with a structure similar to Eq.~(\ref{eq : theta action}). Specifically, we concentrate on third-order terms containing two electromagnetic perturbations and one bosonic perturbation (either a phonon or a magnon). In Fourier space, these terms can be written as
\begin{align}
\label{eq:S3theta}
     &\frac{e^2}{(\beta{\cal V})^3}\text{tr}\sum_{q, p} A_{\mu}(q)A_{\nu}(p)\times\notag\\
     &\left[ i\delta m_5(-q-p)T^{\mu\nu}_5(q, p)+\delta m(-q-p) T^{\mu\nu}(q, p) \right],
\end{align}
with the amplitudes 
\begin{align}
    & T^{\mu\nu}_5(q, p) = \sum_{k}\text{tr}\left[G_0(k)\gamma^{\mu}G_0(k-q)\gamma^{\nu}G_0(k-q-p)\gamma^5\right] \notag\\
    & T^{\mu\nu}(q, p) = \sum_{k}\text{tr}\left[G_0(k)\gamma^{\mu}G_0(k-q)\gamma^{\nu}G_0(k-q-p)\right].
    \notag
  \end{align}
Here, $k$ is the fermion wave (four-)vector, whereas $p$ and $q$ are bosonic wave (four-)vectors. 
The operator $\text{tr}$ denotes the trace over spin and orbital degrees of freedom.
Furthermore, 
\begin{equation}
\label{eq:G0}
    G_0(k) = \frac{-\slashed{k} + m + im_5\gamma^5}{-k^2 + m^2 + m_5^2}
\end{equation}
is the Fourier transform of the free fermion Green's function. We use the convention $-k^2 = \textbf{k}^2 + \kappa_n^2$, with $\kappa_n = \pi(2n+1)/\beta$  and $n\in\mathbb{Z}$.

Using Eq.~(\ref{eq:G0}), applying the identity $\text{tr}(\gamma^{\mu}\gamma^{\nu}\gamma^{\rho}\gamma^{\sigma}\gamma^{5}) = -4\varepsilon^{\mu\nu\rho\sigma}$ \cite{Fujikawa} and assuming a constant magnetic field ${\bf B}_0$ (whereby $\textbf{B}(p) = \beta {\cal V}\textbf{B}_0 \delta_{p,0}$), we identify and collect the axion terms of Eq.~(\ref{eq:S3theta}): 
\begin{equation}
\label{eq:S3theta2}
 S_3^{\theta}=\frac{i e^2}{4\pi^2\beta {\cal V}}\sum_{q}\left[ \delta \theta_{\text{mag}}(q)+\delta\theta_{\rm ph}\right]\textbf{E}(-q)\cdot\textbf{B}_0, 
\end{equation}
where
\begin{align}
\label{eq:dtheta}
    \delta\theta_{\text{mag}}(q) &= -32\pi^2 \mathcal{I}_3(q) m\, \delta m_5(q) \nonumber\\
    \delta\theta_{\text{ph}}(q) &= -32\pi^2 \mathcal{I}_3(q) m_5\, \delta m(q)
\end{align}
are the magnetic and phononic contributions to the dynamical axion, and
\begin{equation}
\label{eq : I3}
   \mathcal{I}_3(q) = \int_k
       \frac{1}{\left(-k^2+\frac{\Delta^2}{4}\right)\left[-(k-q)^2+\frac{\Delta^2}{4}\right]^2}
    	\end{equation}
is an integral with units of energy$^{-2}$ (so that $\delta\theta_{\rm mag}$ and $\delta\theta_{\rm ph}$ are dimensionless).
In Eq.~(\ref{eq : I3}), we have defined
\begin{equation*}
\int_k \equiv \frac{1}{\beta}\sum_{\kappa_n} \int \frac{\di^3 k}{(2\pi)^3}.
\end{equation*}
In Eq.~(\ref{eq:S3theta2}),  the overall $i$ factor ensures that 
the topological part of the action is imaginary in Euclidean signature.

If  $\mathcal{I}_3$ were a constant or weakly dependent on $q$ (which requires a large enough energy gap of the insulator), the axion terms in Eq.~(\ref{eq:dtheta}) could have been easily obtained by applying Fujikawa's method \cite{imaeda2019axion,sekine2021axion, schutte2021axion}. Yet, since we are interested in the $q-$dependence of  $\mathcal{I}_3$, our perturbative approach of computing a triangle Feynman diagram is well-suited.

The dynamical axion $\delta\theta_{\rm mag}$ induced by magnetic fluctuations has been widely discussed (if only in the regime of constant $\mathcal{I}_3$)~\cite{sekine2021axion}.
In contrast, the phononic contribution $\delta\theta_{\rm ph}$ is new. 
It shows that phonons constitute an axion quasiparticle in Dirac insulators that have a nonzero {\em static} $m_5$, without recourse to magnetic fluctuations.

A physical manifestation of $\delta\theta_{\rm ph}$ is the magnetic-field induced phonon effective charge ${\bf Q}_\theta$ introduced in Sec.~\ref{sec:PEC}.
Combining Eqs.~(\ref{eq:Qtheta}) and (\ref{eq:dtheta}), using $\delta m(q) = gu(q)$ and restoring the $\hbar$ factors, we get
\begin{equation}
\label{eq:Qtheta2}
    \textbf{Q}_{\theta}(q) = -8\frac{e^2}{\hbar}m_5 {\cal V}_c g\mathcal{I}_3(q)\textbf{B}_0.
\end{equation}
Let us discuss some salient properties of Eq.~(\ref{eq:Qtheta2}).
First, ${\bf Q}_\theta$ satisfies the expected symmetry conditions for a phonon effective charge: it transforms as a polar vector (because $m_5$ is a pseudoscalar and ${\bf B}_0$ is a pseudovector) and it is even under time-reversal (because both $m_5$ and ${\bf B}_0$ are odd under ${\cal T}$).
Second, ${\bf Q}_\theta$ is a dynamical phonon effective charge, since it depends on the frequency via the integral $\mathcal{I}_3$. 
Third, ${\bf Q}_\theta$ is odd in $m_5$ and even in $m$ because $\mathcal{I}_3$ is an even function of $m_5$ and $m$. 
Consequently, ${\bf Q}_\theta$ can be reversed by reversing the magnetic order that is responsible for $m_5$.

A caveat is in order here. According to our approach, the electronic band parameters ($m$, $m_5$, the Dirac velocity $v$) appearing in Eq.~(\ref{eq:Qtheta2}) are those of bare Dirac electrons. 
Yet, it is known that phonons renormalize those parameters \cite{garate2013phonon, saha2014phonon, kim2015topological, antonius2016temperature, monserrat2016temperature}. 
Thus, a concern might be that the value of $\Delta$ appearing in Eq.~(\ref{eq:Qtheta2}) differs significantly from the experimentally measured energy gap.  
This concern can be assuaged if we reinterpret the band parameters in Eq.~(\ref{eq:Qtheta2}) as being already renormalized by all remaining the phonon modes.
Specifically, suppose that we wish to evaluate the phonon effective charge of a mode $\lambda$ at wave vector $ {\bf q}$. 
We begin by integrating out all the other phonons (i.e. mode $\lambda$ at all ${\bf q}'\neq {\bf q}$, and modes $\lambda'\neq \lambda$ at all ${\bf q}'$.)
Accordingly, the action for the bare Dirac fermions is renormalized. 
Neglecting the frequency-dependence of the electronic self-energy, 
we integrate out the dressed electrons to get an effective action (and from there the effective charge) for the phonon mode $\lambda$ at wave vector ${\bf q}$. The outcome will have the same form as in Eq.~(\ref{eq:Qtheta2}), albeit with renormalized band parameters. With such caveat, in our numerical estimates below we will assume that $\Delta$ corresponds to the experimentally measured value of the energy gap.

The rest of this section will be devoted to a quantitative calculation of ${\bf Q}_\theta$. For experimental expediency, we consider long wavelength optical phonons with $q\simeq ({\bf 0}, \omega_l)$, where $\omega_l = 2\pi l/\beta$  ($l\in\mathbb{Z}$).
An analytical continuation $i\omega_l \to q_0 + i \eta$ (with a small $\eta>0$) allows to determine the physically measured ${\bf Q}_\theta$ as a function of the phonon frequency $q_0$. 
While $\mathcal{I}_3 (q_0)$ is common to magnons and phonons in Eq.~(\ref{eq:dtheta}), earlier works on magnetic axions have only discussed the regimes of zero temperature and $|q_0|\ll \Delta$. Below, we generalize those results to arbitrary gaps and temperatures.



\subsection{Frequency- and temperature-dependence of  ${\bf Q}_\theta$}

\begin{figure*}
   \centering
   \includegraphics[width=\textwidth]{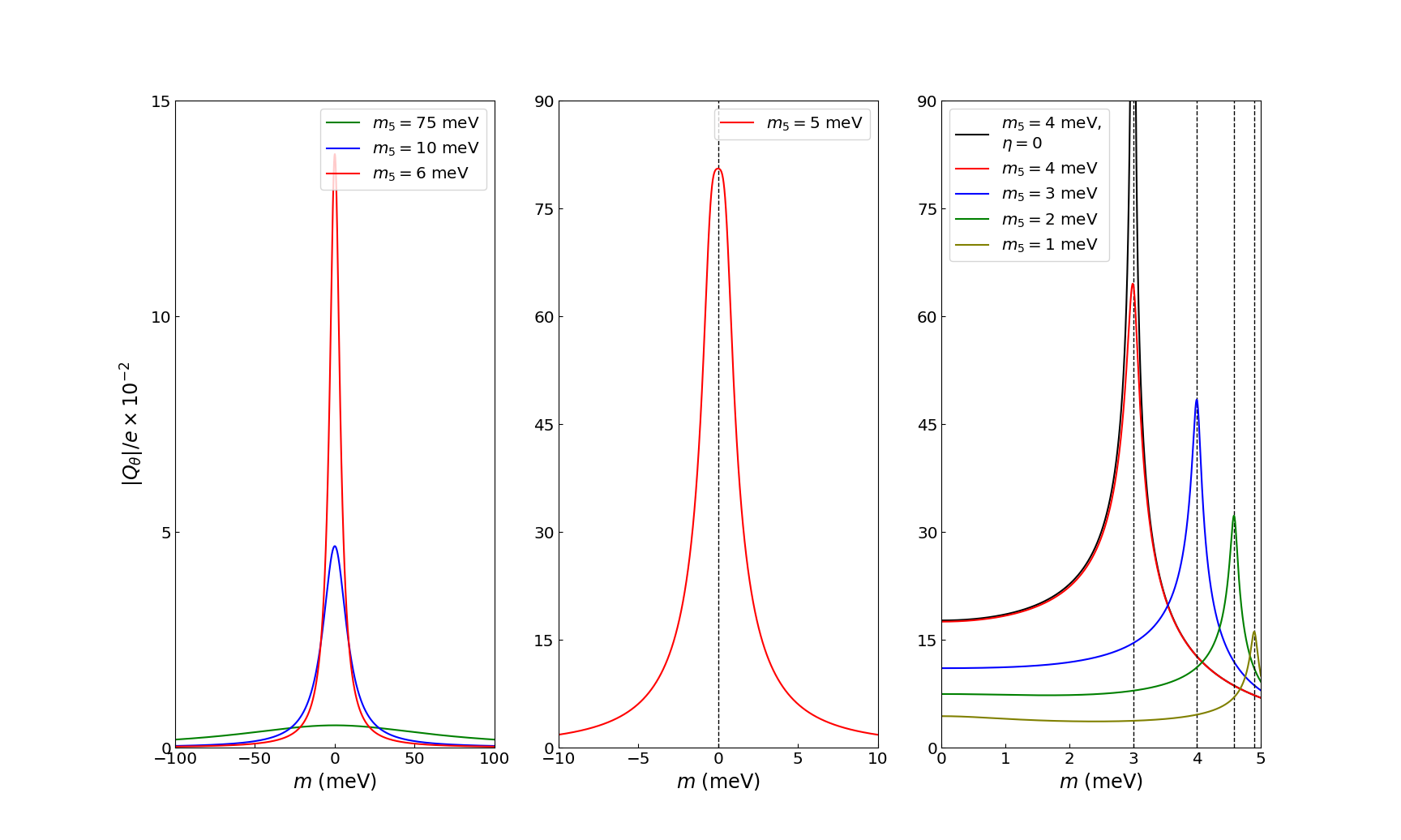}
   \caption{Modulus of the axionic phonon effective charge as a function of the scalar Dirac mass $m$, for fixed axial mass $m_5$.
   Curves were obtained at zero temperature from Eqs.~(\ref{eq:Qtheta2}) and (\ref{eq:zeroT}). 
   The dotted lines represent the values of $m$ for which $\Delta = \hbar q_0$, where $\Delta$ is the energy gap of the insulator and $q_0$ is the phonon frequency.
   Left panel: $|m_5|>\hbar q_0/2$. Middle panel: $|m_5| = \hbar q_0/2$. Right panel: $|m_5|<\hbar q_0/2$.
   The parameter values are $B_0 = 1$T for the magnetic field,  $\hbar q_0 = 10 {\rm meV}$ for the phonon frequency, $g \approx 1 {\rm eV/\AA}$  for the optical deformation potential, $\mathcal{V}_c=1$nm$^3$ for the unit cell volume and $\eta=0.01 \hbar q_0$ for the frequency broadening factor (exception: $\eta = 0$ for the solid black line).}
   \label{fig:Qax_m}
\end{figure*}



The Matsubara sum in Eq.~(\ref{eq : I3}) can be carried out \cite{Nagaosa, Coleman} to find
\begin{align}\label{eq : I3 general}
    \mathcal{I}_3(\omega_l) &= -\frac{1}{16\pi^2}\int_{\Delta/2}^{\infty}\di E\frac{\text{sech}^2\left(\frac{\beta  E}{2}\right)\sqrt{E^2-\Delta^2/4} }{ E^2 \left(4 E^2 +\omega_l^2\right)^2}\times\notag\\
    &\left[\beta  E  \left(4 E^2+\omega_l^2\right)-\sinh (\beta  E) \left(12 E^2+\omega_l^2\right)\right],
\end{align}
for $l \neq 0$. The $l=0$ term can be discarded as we cannot have a zero-frequency optical phonon. 
At zero temperature, Eq.~(\ref{eq : I3 general}) can be solved analytically to obtain
\begin{equation}\label{eq:zeroT}
 \mathcal{I}_3(\omega_l) =  \frac{\ln\left(\omega_l/\Delta+\sqrt{1+\omega_l^2/\Delta^2}\right)}{8\pi^2 \omega_l \Delta\sqrt{1+\omega_l^2/\Delta^2}},
\end{equation}
with $\mathcal{I}_3(q_0) = \mathcal{I}_3(\omega_l\to -i q_0+\eta)$. Let us discuss this result in some detail.

When the bosonic (magnon or phonon) frequency $q_0$ is much smaller than the energy gap $\Delta$ of the insulator, $\mathcal{I}_3(q_0) \simeq 1/(8\pi^2 \Delta^2)$ is approximately constant and real. If in addition $|m_5|\ll |m|$, we have $\mathcal{I}_3(q_0) \simeq 1/(32\pi^2 m^2)$ and therefore recover the well-known result $\delta\theta_{\text{mag}} =-\delta m_5/m$ \cite{sekine2021axion}.

The low-frequency expression $\mathcal{I}_3(q_0) \simeq 1/(8\pi^2 \Delta^2)$ suggests that the dynamical axion fields $\delta\theta_{\rm ph}$ and $\delta\theta_{\rm mag}$  will be largest when $\Delta\to 0$, 
namely at the magnetic and topological phase transition ($m,m_5\to 0$).
Such statement, echoed in the recent literature \cite{zhang2020large, schutte2021axion}, is technically incorrect because
the low-frequency expression for $\mathcal{I}_3(q_0)$ is no longer appropriate when $\Delta$ becomes comparable to or smaller than $\hbar q_0$.
In such regime, $\mathcal{I}_3(q_0)$ is complex and significantly $q_0$-dependent.
In particular, it follows from Eq.~(\ref{eq:zeroT}) that $\mathcal{I}_3(q_0)$, and by association $\delta\theta$ and ${\bf Q}_\theta$, have a maximum not at $\Delta=0$, but instead at $\Delta=\hbar q_0$.
If $\eta\to 0$, this maximum becomes a divergence: for $\Delta/(\hbar q_0)\to 1^+$, the divergence comes from the real part of $\mathcal{I}_3(q_0)$, while for $\Delta/(\hbar q_0)\to 1^-$ the divergence comes from the imaginary part of $\mathcal{I}_3(q_0)$. Therefore $\mathcal{I}_3$ is discontinuous at $\Delta=\hbar q_0$.
A small but finite $\eta$ regularizes the divergence. 
There is also a logarithmic divergence of $\mathcal{I}_3(q_0)$ when $\Delta = 0$, but it is without physical consequence because $\delta\theta_{\rm mag}$ and $\delta\theta_{\text{ph}}$ vanish in this case.

The characteristic energy scale for long-wavelength magnetic fluctuations is small ($q_0\simeq 1 {\rm meV}$) \cite{schutte2021axion}. Therefore, as far as $\delta\theta_{\rm mag}$ is concerned, $\Delta\gg \hbar q_0$ is the experimentally relevant regime for typical values of $\Delta$; this partially justifies the use of a constant and real $\mathcal{I}_3(q_0)$ in earlier studies of magnetic dynamical axions. In contrast, for $\delta\theta_{\rm ph}$, the regime $\Delta\lesssim\hbar q_0$ becomes of increased experimental relevance because the typical optical phonon frequency can easily exceed $10\, {\rm meV}$. 

When $\Delta<\hbar q_0$, the action for the axion quasiparticle is not local in time, i.e. cannot be expressed in the form shown in Eq.~(72) of Ref.~\cite{sekine2021axion}. Indeed, while the bare phonon action is local in time when written in terms of the lattice displacement $u$, it is no longer so when it is 
written in terms of the axion field, as $\delta\theta_{\rm ph}(q_0) \propto  \mathcal{I}_3(q_0) u(q_0)$ implies $\delta\theta_{\rm ph}(t) \propto  \int dt' \mathcal{I}_3(t-t') u(t')$.
A similar statement applies to $\delta\theta_{\rm mag}$.
Thus, the simple axion quasiparticle actions proposed in Refs.~\cite{Li2010,sekine2021axion, schutte2021axion} hold only when the energy gap of the insulator far exceeds the characteristic bosonic (phonon or magnon) frequency and should not be extrapolated to the situation in which the energy gap of the insulator is closing.

For reasonable parameter values ($\hbar q_0\simeq m\simeq m_5\sim 10 {\rm meV}$, $g\simeq 5 {\rm eV}/\AA$, $u\simeq 0.01 \AA$), it follows from Eqs.~(\ref{eq:dtheta}) and (\ref{eq:zeroT}) that $\delta\theta_{\rm ph} \sim 1$.
This estimate suggests that phonon-induced axion terms in Dirac materials can be sizeable at low temperature (in topologically trivial magnetoelectric insulators like BiFeO$_3$ and Cr$_2$O$_3$, the static value of $\theta$ is $\sim 10^{-3}$ \cite{sekine2021axion}).

From Eq.~(\ref{eq:dtheta}), it appears at first glance that $\delta\theta_{\rm ph}$ should vanish when $m_5\to 0$.
A closer inspection shows that to be the case, except when $\hbar q_0 = 2|m|$ and $\eta$ is infinitesimal. Indeed,
\begin{equation}
\label{eq:finetune}
\lim_{m_5\to 0,\eta\to 0} \left(\lim_{\hbar q_0 \to 2 |m|} \delta\theta_{\rm ph}\right)=-g u \frac{\pi}{2} \frac{|m|}{m^2} {\rm sgn}(m_5)
\end{equation}
does not vanish. 
It is remarkable that, when the phonon frequency matches the electronic energy gap, a sizeable phononic axion should be present in materials that break $\mathcal{T}$ and $\mathcal{P}$ only infinitesimally. In practice, a small but nonzero value of $\eta$ and finite temperature $T$ will lead to $\lim_{m_5\to 0} \delta\theta_{\rm ph}=0$. Yet, the remnants of Eq.~(\ref{eq:finetune}) are evident at $\eta, k_B T\ll |m|$, in the form of a pronounced maximum of $\delta\theta_{\rm ph}$ in the vicinity of $m_5= 0$.

\begin{figure*}[t]
   \centering
   \includegraphics[width=\textwidth]{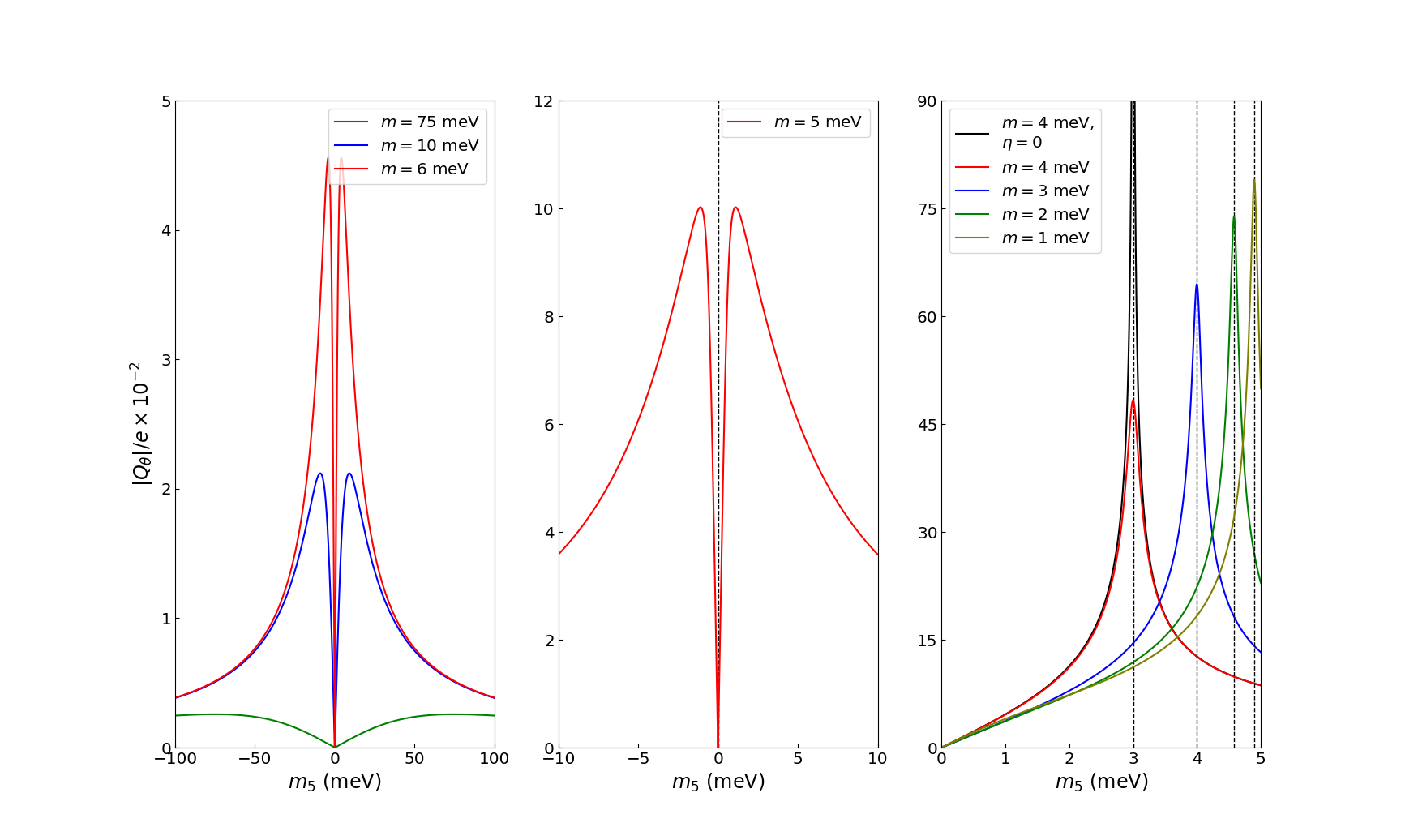}
   \caption{Modulus of the axionic phonon effective charge as a function of the axial mass $m_5$, for fixed scalar mass $m$.
   Curves were obtained at zero temperature from Eqs.~(\ref{eq:Qtheta2}) and (\ref{eq:zeroT}). 
   The dotted lines represent the values of $m_5$ for which $\Delta = \hbar q_0$, where $\Delta$ is the energy gap of the insulator and $q_0$ is the phonon frequency.
   Left panel: $|m|>\hbar q_0/2$. Middle panel: $|m| = \hbar q_0/2$. Right panel: $|m|<\hbar q_0/2$.
   The parameter values are the same as for  Fig.~\ref{fig:Qax_m}.}
   \label{fig:Qax_m5}
\end{figure*}

Figures \ref{fig:Qax_m} and \ref{fig:Qax_m5} illustrate the statements in the preceding paragraphs, for a small but finite value of $\eta$.
In Fig.~\ref{fig:Qax_m}, we show $|{\bf Q}_\theta|$ as a function of $m$, for fixed $m_5$. 
When $2 |m_5|> \hbar q_0$ (i.e. $\Delta>\hbar q_0$ for all values of $m$), $|{\bf Q}_\theta|$ has a maximum at $m=0$, which gets more pronounced for smaller values of $|m_5|$.
The story changes when $2 |m_5|\leq \hbar q_0$. In this case, both $\Delta>\hbar q_0$ and $\Delta<\hbar q_0$ are possible depending on the value of $m$, and a sharp maximum of $|{\bf Q}_\theta|$ is found at the value of $m$ that satisfies $\Delta = \hbar q_0$.

For completeness, Fig.~\ref{fig:Qax_m5} displays $|{\bf Q}_\theta|$ as a function of $m_5$, for fixed $m$.
Similar comments apply as for Fig.~\ref{fig:Qax_m}.
When $2 |m|> \hbar q_0$ (i.e. $\Delta>\hbar q_0$ for all values of $m_5$), $|{\bf Q}_\theta|$ has a maximum for small but nonzero $|m_5|$, which gets more pronounced as $|m|$ gets smaller.
When $2 |m|\leq \hbar q_0$, the maximum of $|{\bf Q}_\theta|$ takes place at $\Delta = \hbar q_0$.
In the special case $2 |m|= \hbar q_0$ (discussed in Eq.~(\ref{eq:finetune})),  $|{\bf Q}_\theta|$ can be large for very small values of $m_5$.

Quantitatively, the preceding figures show that $|{\bf Q}_\theta|\simeq 0.01 e - 0.1 e$ can be attainable for modest magnetic fields (note that our theory is not reliable at high magnetic fields, where Landau quantization of electronic bands should be taken into account).
Such values of $|{\bf Q}_\theta|$ may be experimentally observable, as indicated in the next section.

Thus far, we have concentrated on the zero temperature regime. At finite temperature, Eq.~(\ref{eq : I3 general}) must be solved numerically.  The results for the real and imaginary parts of $Q_\theta$ are displayed in Fig. \ref{fig : Qax(m, T)}.
For simplicity, we have assumed that $m$ and $m_5$ are independent of temperature and that the only $T-$dependence comes from $\mathcal{I}_3$.
When $k_B T\ll\Delta$, we find that the zero-temperature results discussed above still hold to good approximation. When $k_B T\gtrsim \Delta$, however, $\mathcal{I}_3$ and  hence $|{\bf Q}_\theta|$ are strongly suppressed. 
Thus, in order to observe a dynamical axion, it is required that the energy gap of the insulator be large compared to the thermal energy.
The thermal suppression of the axion term has been overlooked or underemphasized in recent references \cite{zhang2020large, schutte2021axion}, which have argued that the largest dynamical axion will take place when $\Delta\simeq 0$.

\begin{figure*}[t]
   \centering
   \includegraphics[width=\textwidth]{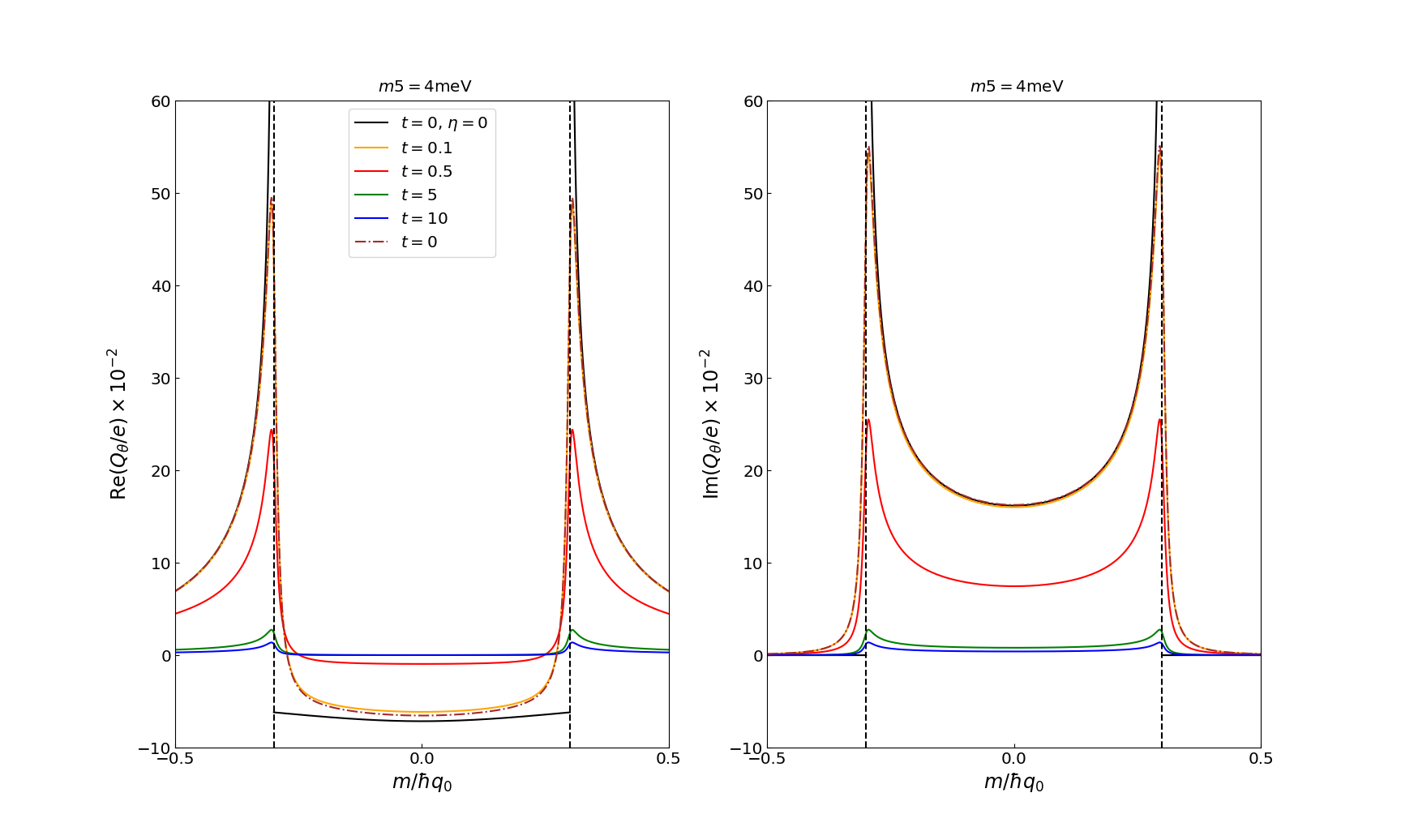}
   \caption{Real and imaginary part of the axionic phonon effective charge as a function of $\Tilde{m} = m/\hbar q_0$ for different values of the dimensionless temperature $t = (\beta\hbar q_0)^{-1}$ and for fixed $m_5 = 4 {\rm meV}$. Curves were obtained from Eqs.~(\ref{eq:Qtheta2}) and (\ref{eq : I3 general}) by numerical integration. The doted lines represent the values of $\Tilde{m}$ for which $\Delta = \hbar q_0$, where $\Delta$ is the energy gap of the insulator and $q_0$ is the phonon frequency ($\hbar q_0 = 10 {\rm meV}$). The parameter values are the same as for Fig.~\ref{fig:Qax_m}.}
   \label{fig : Qax(m, T)}
\end{figure*}

\section{Experimental implications of the phonon-induced dynamical axion}
\label{sec:exp}

In this section, we discuss the physical consequences of the phonon-induced dynamical axion term and the axionic phonon effective charge. 

The phonon-induced axion term modifies Maxwell's equations in a well-established way \cite{qi2008topological, sekine2021axion} and leads to two transport currents of topological origin.
First, the time-derivative of $\delta\theta_{\rm ph}$ produces a chiral magnetic effect, whereby a bulk electric current flows parallel to the magnetic field ${\bf B}_0$.
Second, the space-derivative  of $\delta\theta_{\rm ph}$ leads to an anomalous Hall effect in the presence of an electric field.
For the long-wavelength optical phonons considered in this work, $\delta\theta_{\rm ph}$ is approximately uniform in the bulk of the insulator.
Consequently, the phonon-induced anomalous Hall current is appreciable only at the surface of the insulator, where  $\delta\theta_{\rm ph}$ inevitably has a spatial gradient.

 The preceding phenomena are rather generic to dynamical axions, irrespective of their microscopic origin. 
An alternative, contact-free way to probe the effect of $\delta\theta_{\rm ph}$, which distinguishes the phonon-induced axion from other dynamical axions, is through optical spectroscopy. We discuss this next.

\subsection{Infrared absorption in a magnetic field}

Phonons with nonzero effective charge couple to photons and lead to light absorption. 
The axionic part of the effective charge, ${\bf Q}_\theta$, can be probed by analyzing this absorption.
To quantify this, we start with the Maxwell's equations in Fourier space:
\begin{equation}
\label{eq:max}
-{\bf p} ({\bf p}\cdot{\bf E}) + {\bf p}^2 {\bf E} = \mu_0 q_0^2 (\epsilon_0 {\bf E} + {\bf P}),
\end{equation}
where ${\bf E}$ is the total electric field in the insulator, ${\bf p}$ is its wave vector, $\nu$ is its frequency,
\begin{equation}
\label{eq:pola}
{\bf P} =  {\bf Q} u /{\cal V}_c+ \epsilon_0 \boldsymbol{\chi}_e \cdot {\bf E}
\end{equation}
is the total electronic and lattice polarization, ${\bf Q}={\bf Q}_0+ {\bf Q}_\theta$ is the phonon effective charge
and $\boldsymbol{\chi}_e$ is the electronic susceptibility tensor~\cite{keldysh}.
Herein, we will assume that the Fermi energy lies inside the energy gap of the insulator and that the temperature is much smaller than the gap.
In addition, we will ignore the effect of ${\bf B}_0$ in the electronic susceptibility.
As a result, $\boldsymbol{\chi}_e$ will be approximated as a constant scalar $\chi_e$.

It is convenient to write  ${\bf E} = E_\parallel \hat{\bf p} + {\bf E}_T$, where $E_\parallel \hat{\bf p}$ is the longitudinal electric field produced by lattice vibrations and ${\bf E}_T$ is the transverse electric field due to the incident light (${\bf E}_T\cdot\hat{\bf p} = 0$). 
From Gauss's law, it follows that $\hat{\bf p} \cdot (\epsilon_0 {\bf E} + {\bf P}) = 0$.
Then, Eq.~(\ref{eq:max}) gives
\begin{equation}
\label{eq:max2}
 {\bf p}^2 {\bf E}_T = \mu_0 \nu^2 (\epsilon_0 {\bf E}_T + {\bf P}_T),
 \end{equation}
 where 
${\bf P}_T$ is the transverse part of the polarization.
 
An expression for ${\bf P}_T$ follows from the equation of motion for lattice vibrations~\cite{keldysh,Rinkel2017,Rinkel2019},
\begin{equation}\label{eq : motion phonons}
    M_c(\omega_0^2-(\nu+i \eta)^2)u = \textbf{Q}^*\cdot\textbf{E},
\end{equation}
where $M_c$ is the atomic mass in a unit cell, $\omega_0$ is the bare optical phonon frequency (i.e. excluding the Dirac fermion contribution to it) and ${\bf Q}^*(\nu)={\bf Q}(-\nu)$.
Combining Gauss' law with Eq.~(\ref{eq:pola}), Eq.~(\ref{eq : motion phonons}) can be rewritten as
\begin{equation}
\label{eq:mot2}
   M_c\left(q_0^2-(\nu+i \eta)^2\right)u = \textbf{Q}^*\cdot\textbf{E}_T,
\end{equation}
where
\begin{equation}
\label{eq:lo}
q_0^2 \equiv \omega_0^2+\frac{|{\bf Q}\cdot\hat{\bf p}|^2}{M_c \mathcal{V}_c \epsilon_0 (1+\chi_e)}.
\end{equation}
As a result, we arrive at
\begin{equation}
\label{eq:pt}
{\bf P}_T =  {\bf Q}_T \frac{{\bf Q}^*\cdot{\bf E}_T}{M_c\mathcal{V}_c(q_0^2-(\nu+i \eta)^2)}+\epsilon_0 \chi_e {\bf E}_T,
\end{equation}
where ${\bf Q}_T \equiv {\bf Q}- \hat{\bf p} ({\bf Q}\cdot\hat{\bf p})$ is the transverse part of the phonon effective charge.

Substituting Eq.~(\ref{eq:pt}) in Eq.~(\ref{eq:max2}) and solving the resulting system of equations, one can obtain the dispersion and the attenuation of light waves entering the crystal.
Let us discuss some simple cases of interest, for which ${\bf Q}$ is either parallel or perpendicular to the photon wave vector ${\bf p}$.
When ${\bf Q} || \hat{\bf p}$ (longitudinal optical phonon), Eq.~(\ref{eq:max2}) results in ${\bf p}^2/\nu^2 = \mu_0 \epsilon_0 (1+\chi_e)$. 
Thus, in this case there is no phonon signature in light absorption.
In the case of a transverse optical phonon (${\bf Q}\perp \hat{\bf p}$), two solutions of Eq.~(\ref{eq:max2}) are possible. 
In the first solution, $ {\bf E}_T\perp {\bf Q}$ and ${\bf p}^2/\nu^2 = \mu_0 \epsilon_0 (1+\chi_e)$; thus, the phonon remains invisible in light absorption spectroscopy. 
In the second solution, $ {\bf E}_T || {\bf Q}$ and  ${\bf p}^2/\nu^2 = \mu_0 \epsilon$, with a dielectric function
\begin{equation}
\label{eq:epsilon}
\epsilon = \epsilon_0 (1+\chi_e) + \frac{|{\bf Q}|^2}{M_c \mathcal{V}_c (\omega_0^2-(\nu+i \eta)^2)}.
\end{equation}
In Eq.~(\ref{eq:epsilon}), the phonon contributes to $\epsilon$ and thus to light absorption.
Specifically, the absorption coefficient (in units of inverse length)  is given by \cite{cardona2005fundamentals} $\alpha = \nu  \chi_I/( n c)$, where $n$ is the refractive index, $c$ is the speed of light in vacuum and $\chi_I$ is the imaginary part of $\epsilon/\epsilon_0$.

\begin{figure}[t]
\centering
\includegraphics[width=1\linewidth]{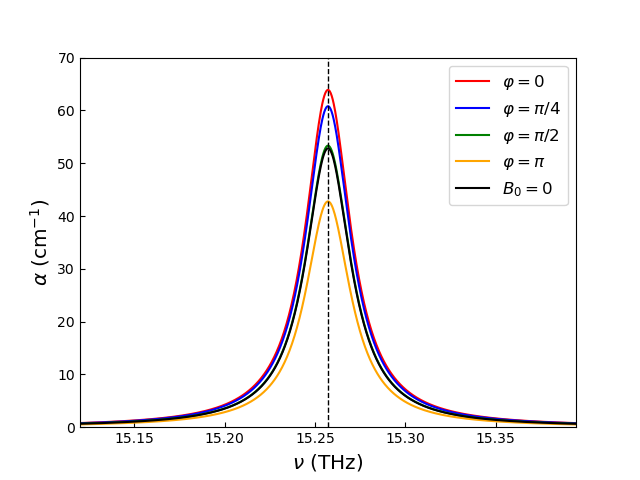}
\caption{Phonon contribution to the light absorption coefficient at zero temperature, as a function of the photon frequency $\nu$. Curves were calculated from Eq.~(\ref{eq:epsilon}). The doted line corresponds to $\nu=\omega_0$, where $\omega_0$ is the transverse (and infrared active) optical phonon frequency. The variable $\varphi$ represents the angle between the conventional phonon effective charge ($\textbf{Q}_0$) and the external magnetic field ($\textbf{B}_0$). The parameter values are $B_0 = 1$T (with the exception of a curve for which $B_0=0$), $Q_0 = e$ for the conventional phonon effective charge, $Q_{\theta} = 0.1e$ for the axionic phonon effective charge, and $\eta=0.01 \omega_0$ for the frequency broadening factor.}
\label{fig : absorption}
\end{figure}

The phonon contribution to light absorption is shown in Fig.~\ref{fig : absorption}.
It illustrates the most favorable scenario for probing ${\bf Q}_\theta$ in infrared absorption, which is when ${\bf Q}_0$ is nonzero and sizeable (of the order of $e$).
Generally, ${\bf Q}_0$ has a fixed direction in space dictated by the crystal and the phonon mode symmetry, and its magnitude is independent from (or weakly dependent on) the magnetic field ${\bf B}_0$. 
In contrast, ${\bf Q}_\theta$ is proportional and parallel to ${\bf B}_0$.
In order to simplify our discussion and use Eq.~(\ref{eq:epsilon}), we suppose that both ${\bf Q}_0$ and ${\bf B}_0$ are in the plane perpendicular to ${\bf p}$.
When $B_0=0$, only ${\bf Q}_0$ contributes to the infrared absorption. 
As $B_0$ is turned on, ${\bf Q}_\theta$ influences the infrared absorption intensity.
Since $|{\bf Q}_0| \gg |{\bf Q}_\theta|$, the leading axionic contribution to $\epsilon$ in Eq.~(\ref{eq:epsilon}) goes like ${\bf Q}_0\cdot {\bf Q}_\theta$, whose sign is reversed by reversing ${\bf B}_0$.
In passing, this is the reason why a large ${\bf Q}_0$ is helpful for the detection of $|{\bf Q}_\theta|$. 
Should ${\bf Q}_0$ vanish (which would be the case if the phonon of interest were infrared inactive at zero magnetic field), then the axionic correction to $\epsilon$ would scale as $|{\bf Q}_\theta|^2$, which is numerically small compared to $|{\bf Q}_0| |{\bf Q}_\theta|$.
Figure~\ref{fig : absorption} displays variations in the absorption coefficient as ${\bf B}_0$ is rotated. 
For $|{\bf Q}_0|\simeq e \simeq 10 |{\bf Q}_\theta|$, the variations in $\alpha$ exceed $10\, {\rm cm}^{-1}$ and are thus in principle large enough to be measurable in ellipsometry \cite{cardona2005fundamentals}.

In experiment, electrons will also contribute to the absorption coefficient $\alpha$ through the imaginary part of $\chi_e$ in Eq.~(\ref{eq:epsilon}).
To avoid masking the phonon contribution discussed above,
insulating materials with Fermi level inside the bulk gap and temperature smaller than the gap are desirable. 

\subsection{LO-TO splitting}

As evidenced by Eqs.~(\ref{eq:mot2}) and (\ref{eq:lo}), longitudinal and transverse optical phonons have different frequencies. 
Assuming that the difference in frequencies ($\Delta_{\rm LO-TO}$) is small compared to their sums, we can write
\begin{equation}
\Delta_{\rm LO-TO} \simeq \frac{|{\bf Q}|^2}{2 \omega_0 M_c \mathcal{V}_c \epsilon_0 (1+\chi_e)},
\end{equation}
where $\omega_0$ is the frequency of the transverse optical (TO) phonon, and the dynamical part of the effective charge Q can be evaluated at frequency $\omega_0$ to good approximation \footnote{Solving Eq.~(\ref{eq:lo}) numerically with input from Eqs.~(\ref{eq:Qtheta2}) and (\ref{eq:zeroT}), we find that the frequency $q_0$ of the longitudinal optical (LO) phonon is very close to $\omega_0$ even when $Q_\theta$ is maximal (at $\Delta/(\hbar q_0) \simeq 1$).}.

Based on the numerical estimates of the preceding section, we will assume that $Q_0\gg Q_\theta$.
When $Q$ changes by a quantity $\delta Q$ (e.g. due to a change in the magnetic field), the LO-TO splitting changes by
\begin{equation}
\delta\Delta_{\rm LO-TO}\sim \frac{Q_0 \delta Q}{q_0 M_c \mathcal{V}_c \epsilon_0 (1+\chi_e)}.
\end{equation}
Assuming an experimental resolution of $\sim 0.1 {\rm cm}^{-1} \sim 0.01 {\rm meV}$ in $\Delta_{\rm LO-TO}$,
we can extract the minimum value of $\delta Q$ (and thus $Q_\theta$) that can be experimentally resolved.
For reasonable parameters ($\chi_e\simeq 20$, $\mathcal{V}_c\simeq 125 \AA^3$, $M_c\simeq 10^{-24} {\rm kg}$, $\hbar q_0 \simeq 10 {\rm meV}$, $Q_0 \simeq e$), it follows that $\delta Q$ (and thus $Q_\theta$) must exceed $\sim 0.2 e$ in order to be observable. 
It is not inconceivable to have such values of $Q_\theta$, though detection appears challenging and may require high magnetic fields.




\section{Discussion and conclusions}
\label{sec:disc}
In summary, we have presented a theoretical study on how lattice vibrations can produce a dynamical axion in three dimensional Dirac insulators with 
a static axial mass $m_5$. The phononic axion manifests physically through a phonon effective charge (the ``topological" or ``axionic''  PEC) that is proportional and parallel to an external static magnetic field.
This axionic PEC switches sign under $m_5 \to -m_5$ (when the magnetic order responsible for $m_5$ is reversed), and is maximal when the electronic energy gap coincides with the phonon frequency.
Our work hopes to increase the observability of the axion quasiparticles in insulators by proposing additional, phonon-based,  probes for their detection and excitation.

Phonon-induced axion terms were predicted earlier in Weyl semimetals \cite{Rinkel2017, Rinkel2019}, and their possible observation reported in experiment \cite{Yuan2020}. 
Our present study can be regarded as an extension of those earlier works to insulating materials. 
The main similarity between topological semimetals and insulators resides in the general form of the axionic PEC (Eq.~(\ref{eq : PEC topo}) in the present work), which is the same for both systems. There are nonetheless quantitative differences. In Weyl semimetals, the axionic phonons are those that couple to electrons as axial gauge fields. 
In contrast, in Dirac insulators with broken time-reversal and space-inversion symmetries, scalar phonons can also behave like axion quasiparticles.
In addition, the frequency and temperature-dependence of the phonon-induced axion term is very different in semimetals and insulators, owing to the energy gaps of the latter.

Candidate insulators in which our results can be tested are primarily those belonging to the MnBiTe family \cite{wang2021}.
These are layered materials with intrinsic magnetic order. In a given layer, the magnetic moments of the Mn atoms are aligned with one another in the direction perpendicular to the layer. At the same time, the Mn magnetic moments in adjacent layers are antialigned with respect to one another.
Mn$_2$Bi$_2$Te$_4$ thin films with an even number of septuple layers \cite{zhu2021tunable} and bulk Mn$_2$Bi$_2$Te$_5$ \cite{li2020,zhang2020large}  have been theoretically predicted to host Dirac fermions with $m_5\neq 0$
\footnote{These first-principles studies do not consider the zero-point renormalization effects for $m$ and $m_5$ originating from electron-phonon interactions, which could be significant and could be studied following the approaches outlined in e.g.  Ref. \cite{miglio2020predominance}.}.
Following conventional wisdom, the dynamical axion in these materials has been attributed to fluctuations in the magnetic order.
Yet, in the present paper we have clarified that ubiquitous lattice vibrations can themselves lead to a dynamical axion even when the magnetic order does not fluctuate.
While the coupling between the magnetic order in MnBi$_2$Te$_4$ and the phonons has been recently measured \cite{padmanabhan2022}, no connection has been recognized or explored between phonons and axions.

One subtlety of Mn$_2$Bi$_2$Te$_5$ is that it is near a ferromagnetic-antiferromagnetic phase boundary \cite{tang2023intrinsic}. As a result, the application of a modest magnetic field of $\lesssim 10$ T \cite{lee2021} can induce a phase transition from the antiferromagnetic to the ferromagnetic phase \cite{li2020}, the latter having $m_5=0$. 
As such, the topological PEC will be a nonmonotonic function of the magnetic field in this material.
We note also that the temperature-dependence of $m_5$ (neglected in our work) can be taken advantage of for the purposes of experimental detection of the topological PEC. 
For example, $m_5$ (and consequently the topological PEC)  will vanish above the N\'eel transition temperature.

In order to avoid complications arising from the magnetic-field- and temperature-dependence of $m_5$, it would be more convenient to have a material with stronger antiferromagnetic order. 
Recently, Ni$_2$Bi$_2$Te$_5$ has been theoretically predicted \cite{tang2023intrinsic} to be a dynamical axion insulator with a N\'eel temperature approaching room temperature (as opposed to $~20$K in Mn$_2$Bi$_2$Te$_5$).
In this case, the ground state magnetic order (i.e. the value of $m_5$)  is presumably more robust under the application of a magnetic field, and a field-induced topological PEC can be more easily observed.

There are various possible directions for future research. 
On the technical side, our theory could be refined by including certain aspects that have been omitted, such as the Zeeman splitting, Landau levels and electron-electron interactions. 
On the conceptual side, our theory could be extended by incorporating surface effects. 
In this paper, we have for simplicity considered an infinite bulk. It could be interesting to calculate the phonon-induced axion dynamics when the bulk preserves inversion and time-reversal symmetries (so that $\theta$ is quantized) but the surface breaks both symmetries. 
In this situation, following the heuristic arguments of Sec.~\ref{sec:PECax}, the phonon effective charge might be a diagnostic tool for a topological phase transition.
On the practical side, our predictions are based on a toy model and the numerical parameter values adopted therein (the energy gap of the insulator, the phonon frequency, the unit cell volume, the strength of the electron-phonon coupling, etc.) are for indicative purpose only. 
Since the observability of the effects we predict is contingent on said values, it appears important to adapt our theory to real materials with input from first-principles calculations.  

Finally, it could  be interesting to explore the implications of our theory for the ongoing search of the cosmological axion.
Recent papers have proposed to detect dark matter using phonons with conventional effective charges in nontopological materials \cite{mitridate2020, marsh2023axion}, or magnetic axions in topological materials \cite{marsh2019}. 
A natural question is whether the phononic axion quasiparticles proposed in our work, with their magnetically tunable phonon effective charge, could have any use in the detection of the cosmological axion.

\acknowledgements
This work has been financially supported by the Canada First Research Excellence Fund, the Natural Sciences and Engineering Research Council of Canada (Grant No. RGPIN- 2018-05385), and the Fonds de Recherche du Qu\'ebec Nature et Technologies.
We thank B. Groleau-Par\'e for technical assistance in the initial stages of the project, and L. Fu for a helpful discussion.

\bibliography{bibliography}{}

\end{document}